\documentclass[a4paper,11pt]{article}
\pdfoutput=1

\usepackage{jcappub} 
\usepackage{natbib}
\setcitestyle{square,comma,numbers,sort&compress}
\usepackage{enumerate}
\usepackage{tabularx}
\usepackage{bbold}
\usepackage[left = 1.0in, top=1.0in,right=1.0in,bottom=1.0in,a4paper]{geometry}
\usepackage[dvipsnames]{xcolor}
\usepackage{comment}
\usepackage[caption=false]{subfig}
\usepackage{cancel}

\usepackage[section]{placeins}
\usepackage{listings}
\usepackage{amsbsy}
\usepackage{empheq}
\newcolumntype{Y}{>{\centering\arraybackslash}X}

\definecolor{myRED}{rgb}{0.8, 0.25, 0.33}
\usepackage[normalem]{ulem}
\usepackage{dsfont}
\usepackage{pbox}
\usepackage{graphicx}
\usepackage{multirow}
\usepackage{tikz}
\usetikzlibrary{arrows}
\usepackage{enumitem} 
\usepackage{ulem}
\usepackage{soul}
\usepackage{lipsum}  
\usepackage{changepage}

\allowdisplaybreaks
\title{\boldmath\LARGE  Updated Running Quark and Lepton Parameters \\at Various Scales}
\author[a]{Stefan Antusch,}
\author[a]{Kevin Hinze,}
\author[b]{Shaikh Saad}

\affiliation[a]{Department of Physics, University of Basel, Klingelbergstrasse\ 82, CH-4056 Basel, \\Switzerland}

\affiliation[b]{Jožef Stefan Institute, Jamova 39, P.\ O.\ Box 3000, SI-1001 Ljubljana, Slovenia}

\emailAdd{stefan.antusch@unibas.ch, kevin.hinze@unibas.ch, shaikh.saad@ijs.si}
\abstract{
In the light of the recent Particle Data Group (PDG) release, we revisit the running quark and lepton Yukawa couplings, together with the quark mixing parameters, across a range of energy scales. The 2024 PDG determinations of low-energy fermion masses feature significantly smaller  uncertainties, resulting from a reduced estimate of systematic errors compared to the more conservative treatment in the 2022 analysis. To assess the impact of these changes, we present running parameters obtained using both the 2022 and 2024 datasets, within the frameworks of the Standard Model (SM) and its minimal supersymmetric extension  (MSSM). The evolved values, along with their associated $1\sigma$ uncertainties, are given within the SM framework at benchmark scales of $M_Z$ and $10^3$, $3\cdot 10^3$, $10^4$, $10^5$, $10^7$, $10^9$, $10^{12}$, and $10^{16}$ GeV. Within the MSSM, we additionally provide GUT-scale results for different choices of $\tan\beta$, assuming supersymmetry breaking scales of 3 and 10 TeV, including an approximate way for taking supersymmetric loop threshold corrections into account. We furthermore discuss implications of the updated results for constructing and testing theories beyond the SM. 
}
\makeatletter
\gdef\@fpheader{}
\makeatother
\begin{document}
\maketitle
\flushbottom

\section{Introduction}
The origin of quark and lepton masses and their strongly hierarchical structure remains one of the most profound unresolved puzzles in particle physics. Although numerous theoretical frameworks have been proposed to account for the observed patterns of masses and mixings, a definitive explanation has yet to be established. In this context, the precise determination of the running quark and lepton masses, together with the mixing parameters, are of central importance, as they provide stringent constraints on models of fermion masses and flavour dynamics. In fact, our knowledge of fermion masses and mixing parameters has improved substantially in recent years, leading to increasingly precise determinations of these quantities. In particular, the most recent Particle Data Group (PDG) release in 2024 reports a significant reduction in the uncertainties of quark masses compared to earlier analyses, such as the 2022 PDG values. Motivated by these developments, it is timely to revisit the running quark and lepton Yukawa couplings, along with the quark mixing parameters, which serve as essential input for model building efforts in the flavour sector. Over the last 30 years, numerous systematic calculations of the running parameters across various energy scales have been considered, e.g., Refs.~\cite{Fusaoka:1998vc,Das:2000uk,Ross:2007az,Xing:2007fb,Xing:2011aa,Antusch:2013jca,Deppisch:2018flu,Huang:2020hdv}, which have proven very useful for model-building studies.

For model-building purposes, it is helpful to have access to the running fermion masses and mixing parameters at the high-energy scale where the theory is defined. Alternatively, having these quantities at a reference low-energy-scale -- such as the $Z$ boson mass, $M_Z$ -- allows for a direct comparison between model predictions  and experimental data after performing the renormalization group (RG) evolution  from the high-energy-scale down to the $M_Z$-scale. Therefore, in this work, in light of the recent PDG releases, we revisit the running  quark and lepton Yukawa couplings, together with the quark mixing parameters, over a wide range of energy scales.
We present running parameters derived from both the 2022 and 2024 PDG datasets, within the frameworks of the Standard Model (SM) and its minimal  supersymmetric extension (MSSM). The evolved values, along with their associated $1\sigma$ uncertainties, are provided within the SM framework at benchmark scales  of $M_Z$, $10^3$, $3\times10^3$, $10^4$, $10^5$, $10^7$, $10^9$, $10^{12}$,  and $10^{16}$ GeV.  For the MSSM, we furthermore report the values at the unification scale of Grand Unified Theories (GUTs), i.e.\ the GUT-scale (taken to be $M_\mathrm{GUT}=2\times 10^{16}$ GeV), for several representative values of $\tan\beta$, namely $\tan\beta = 5$, 10, 30, and 50, assuming supersymmetry-breaking-scales of $M_\mathrm{SUSY}=$ 3 and 10 TeV. 
It is important to emphasize that, in the SUSY case, our results are presented for the case of zero SUSY-scale loop threshold corrections, which can be $\tan \beta$-enhanced and thus play an important role in the down-type quark and charged lepton Yukawa sector. To take them into account, we provide an approximate method for implementing them at the GUT-scale, applicable to cases where $\tan\beta$ is not too large and/or the threshold correction parameters are not too big.

The manuscript is organized in the following way. In section~\ref{sec:02}, we first discuss the procedure for obtaining the $M_Z$-scale values of the fermion masses and mixing parameters from low-energy experimental inputs. These quantities, together with their corresponding uncertainties, are then evolved in the $\overline{\text{MS}}$ renormalization scheme to various energy scales within the SM framework. The corresponding results are summarized in Tables~\ref{MS, 2022}-\ref{MS, 2024}. Section~\ref{sec:03} focuses on obtaining the GUT-scale values of these parameters in the context of the MSSM. For this purpose, the converted values in the $\overline{\text{DR}}$ renormalization scheme are tabulated in Tables~\ref{DR, 2022}-\ref{DR, 2024}. Finally, the resulting running values at the GUT-scale are presented in Tables~\ref{2022, GUT, 3TeV}-\ref{2024, GUT, 3TeV} and Tables~\ref{2022, GUT, 10TeV}-\ref{2024, GUT, 10TeV} for $M_\textrm{SUSY} = 3$ TeV and $M_\textrm{SUSY} = 10$ TeV, respectively. Furthermore, in section~\ref{sec:04}, we discuss implications of the updated results for constructing and testing theories beyond the SM. 
Finally, we conclude in section~\ref{sec:con}.

\section{SM running parameters}\label{sec:02}
Let us first discuss the 15 low-scale input parameters. These are the top quark pole mass $M_t$, the Higgs boson pole mass $M_h$, three charged lepton pole masses  $M_\tau$, $M_\mu$, $M_e$, and the $Z$ boson Breit-Wigner mass $M_Z^\text{BW}$. Additionally, there are five light quark running masses $m_b(m_b)$, $m_c(m_c)$, $m_s(2\,\text{GeV})$, $m_d(2\,\text{GeV})$, $m_u(2\,\text{GeV})$, the Fermi constant $G_F$, the 5-quark QCD coupling  $\alpha_s^{(5)}(M_Z)$  ($\alpha_s^{(5)}$  is the strong coupling constant $\alpha_s$ with 5 flavours  of quarks, $n_f=5$),  the fine structure constant $\alpha_0$ and the hadronic correction parameter $\Delta\alpha_\text{had}^{(5)}(M_Z)$. Values of these parameters according to the 
\textbf{2022 PDG} data\ \cite{ParticleDataGroup:2022pth} are  
\begin{align}
    &M_t=172.5\pm0.7\;\text{GeV},
    \quad
    M_h=125.25\pm 0.17\;\text{GeV},
    \quad
    M_Z^\textrm{BW}=91.1876\pm 0.0021\;\text{GeV},
    \nonumber\\
    &m_b(m_b)=4.18^{+0.03}_{-0.02}\;\text{GeV},
    \quad
    m_c(m_c)=1.27\pm 0.02\;\text{GeV},
    \quad
    m_s(2\,\text{GeV)}=93.4^{+8.6}_{-3.4}\;\text{MeV},
    \nonumber\\
    &m_d(2\,\text{GeV)}=4.67^{+0.48}_{-0.17}\;\text{MeV},
    \quad
    m_u(2\,\text{GeV)}=2.16^{+0.49}_{-0.26}\;\text{MeV},
    \quad
    M_\tau=1776.86\pm 0.12\;\text{MeV},
    \nonumber\\
    &
    M_\mu=105.6583755\pm 0.0000023\;\text{MeV},
    \quad
    M_e=0.51099895000\pm 0.00000000015\;\text{MeV},
    \nonumber\\
    &G_F=(1.1663788\pm 0.0000006)\times 10^{-5}\;\text{GeV}^2,
    \quad
    \alpha_0^{-1}=137.035999180\pm0.000000010,
    \nonumber\\
    &\alpha_s^{(5)}(M_Z)=0.1179\pm 0.0009,
    \quad 
    \Delta\alpha_\text{had}^{(5)}(M_Z)=0.02768\pm 0.00007 \,. \label{eq:2022}
\end{align}
The values of these parameters according to the \textbf{2024 PDG} data\ \cite{ParticleDataGroup:2024cfk} are\footnote{ For all quark masses except that of the top quark pole mass, which are reported at 90\% confidence level, the corresponding errors are rescaled to their $1\sigma$ values. }
\begin{align}
    &M_t=172.4\pm 0.7\;\text{GeV},
    \quad
    M_h=125.20\pm 0.11\;\text{GeV},
    \quad
    M_Z^\textrm{BW}=91.1880\pm 0.0020 \;\text{GeV},
    \nonumber\\
    &m_b(m_b)=4.183\pm 0.004\;\text{GeV},
    \;
    m_c(m_c)=1.2730\pm 0.0028\;\text{GeV},
    \;
    m_s(2\,\text{GeV)}=93.5\pm 0.5\;\text{MeV},
    \nonumber\\
    &m_d(2\,\text{GeV)}=4.70\pm 0.04\;\text{MeV},
    \;\;
    m_u(2\,\text{GeV)}=2.16\pm 0.04\;\text{MeV},
    \;\;
    M_\tau=1776.93\pm 0.09\;\text{MeV},
    \nonumber\\
    &
    M_\mu=105.6583755\pm 0.0000023\;\text{MeV},
    \quad
    M_e=0.51099895000\pm 0.00000000015\;\text{MeV},
    \nonumber\\
    &G_F=(1.1663788\pm 0.0000006)\times 10^{-5}\;\text{GeV}^2,
    \quad
    \alpha_0^{-1}=137.035999178\pm 0.000000008,
    \nonumber\\
    &\alpha_s^{(5)}(M_Z)=0.1180\pm 0.0009,
    \quad 
    \Delta\alpha_\text{had}^{(5)}(M_Z)=0.02783 \pm 0.00006 \,. \label{eq:2024}
\end{align}

In this work, we update the analysis of Ref.~\cite{Antusch:2013jca} to obtain the running parameters using both the 2022 and 2024 PDG releases.
Looking at these inputs, Eqs.~\eqref{eq:2022} and \eqref{eq:2024}, one can immediately notice that the quark masses, according to the 2024 PDG analysis, are provided with unprecedented accuracy, far surpassing the precision previously stated. For example, a comparison of the 2022 and 2024 PDG quoted values shows that, for the bottom quark, the uncertainty has decreased from 0.7\% to 0.09\%, and similarly for the other quark masses.  On the other hand, the central values of the quark and lepton masses do not change significantly between the 2022 and 2024 PDG releases. A quick comparison between Eqs.~\eqref{eq:2022} and \eqref{eq:2024} immediately reveals that the largest shift occurs in the down-quark mass; however, this amounts to only a 0.6\% change. Thus, the primary difference instead lies in the quoted uncertainties: the 2022 edition adopted a more conservative assessment of certain systematic effects, whereas the 2024 release reflects increased confidence in controlling these uncertainties and therefore reports smaller errors. In this context, presenting results obtained with both PDG releases serves as a sensitivity analysis with respect to variations in the fundamental input parameters (which remain largely stable between 2022 and 2024) and their associated uncertainties (which are reduced significantly in 2024). Displaying both sets of results enables model builders to assess how sensitive their model parameters are to the different treatments of uncertainties in the fundamental input quantities.

To compute the RG running of these input parameters up to the $M_Z$-scale and to convert them into the parameters of the SM ($y_u$, $y_d$, $y_s$, $y_c$, $y_b$, $y_t$, $y_e$, $y_\mu$, $y_\tau$, $g_1$, $g_2$, $g_3$, $\lambda$, and $v$) in the $\overline{\text{MS}}$ renormalization scheme\footnote{Note that, compared to Ref.\ \cite{Martin:2019lqd}, we use a different normalization for the Higgs quartic coupling $\lambda$. Namely, we define the Higgs potential as $V(\phi)=m^2|\phi|^2+\tfrac{1}{4}\lambda|\phi|^4$. Moreover, we use the GUT normalization for the gauge coupling $g_1$, i.e.\ $g_1^2=5/3g^{\prime 2}$.}, 
we use \texttt{SMDR}\ \cite{Martin:2019lqd}, a software library written in C. 
In addition to the QCD contributions to the RG running and matching, which are implemented in the \texttt{Mathematica} code \texttt{RunDec}\ \cite{Chetyrkin:2000yt, Herren:2017osy} (and C code \texttt{CRunDec}\ \cite{Schmidt:2012az, Herren:2017osy}) and were used in Ref.\ \cite{Antusch:2013jca}, \texttt{SMDR} also handles the 2-loop pure electroweak (EW) and mixed QCD/EW  contributions to the running and matching.

Now, using the low-scale PDG inputs and sampling over 100,000 points with assumed Gaussian errors, we obtain the following SM parameters along with their $1\sigma$ uncertainties, namely, the highest posterior density (HPD) intervals, at the $M_Z$-scale in the $\overline{\text{MS}}$ renormalization scheme. For \textbf{2022 PDG} input we find 
\begin{align}\label{eq:MZ2022}
    &g_1=0.461227^{+0.000025}_{-0.000027}\,,
    &&g_2=0.65096^{+0.00004}_{-0.00004}\,,
    &&g_3=1.2118^{+0.0045}_{-0.0047}\ ,\nonumber\\
    &y_u=(7.09^{+1.56}_{-0.88}){\times} 10^{-6}, 
    &&y_d=(1.55^{+0.14}_{-0.07}){\times} 10^{-5},
    &&y_s=(3.10^{+0.26}_{-0.14}){\times} 10^{-4},\nonumber\\
    &y_c=(3.55^{+0.10}_{-0.09}){\times} 10^{-3},
    &&y_b=(1.63^{+0.02}_{-0.01}){\times} 10^{-2},
    &&y_t=0.968^{+0.004}_{-0.004}\,,\nonumber\\
    &y_e=(2.77705^{+0.00033}_{-0.00039}){\times} 10^{-6}, 
    &&y_\mu=(5.85026^{+0.00076}_{-0.00075}){\times} 10^{-4}, 
    &&y_\tau=(0.99370^{+0.00015}_{-0.00014}){\times} 10^{-2},\nonumber\\
    &\lambda=0.5593^{+0.0018}_{-0.0020}\,,
    &&v=248.408^{+0.028}_{-0.036}\;\text{GeV}.
\end{align}
Note that for the Higgs vacuum expectation value at $\mu = M_t$ we get $v=246.603$\;GeV. 

For \textbf{2024 PDG} input the results are
\begin{align} \label{eq:MZ2024}
     &g_1=0.461228\pm0.000026\,,
     &&g_2=0.65096\pm0.00004\,,
     &&g_3=1.2123\pm0.0046\,,\nonumber\\
     &y_u=(7.04\pm0.15){\times} 10^{-6},
     &&y_d=(1.54\pm0.02){\times} 10^{-5},
     &&y_s=(3.06\pm0.04){\times} 10^{-4},\nonumber\\
     &y_c=(3.56\pm0.06){\times} 10^{-3},
     &&y_b=(1.630\pm0.009){\times} 10^{-2},
     &&y_t=0.967\pm0.004\,,\nonumber\\
     &y_e=(2.77713\pm0.00036){\times} 10^{-6},
     &&y_\mu=(5.85042\pm 0.00075){\times} 10^{-4},\nonumber\\
     &y_\tau=(0.99378\pm0.00014){\times} 10^{-2},
     &&\lambda=0.55853\pm0.00157\,,
     &&v=248.401\pm0.032\;\text{GeV}.
\end{align} 
The Higgs vacuum expectation value at $\mu = M_t$ is now given by $v=246.604$\;GeV.

For the Cabibbo–Kobayashi–Maskawa (CKM) mixing parameters we use the updated UTfit 2023 results\ \cite{UTfit:2022hsi} as input: 
\begin{align}
    &\sin\theta_{12}=0.2251\pm0.0008\,,
    &&
    \sin\theta_{23}=0.04193\pm0.00041\,,
    \nonumber\\
    &
    \sin\theta_{13}=0.00370\pm0.00008\,,
    && 
    \delta=1.139\pm0.023 \,.
\end{align}
They correspond to the following values of the CKM mixing angles:
\begin{align}
     &\theta_{12}=0.2270\pm 0.0008\,,
     &&\theta_{23}=(4.194\pm{0.041})\times 10^{-2},
     &&\theta_{13}=(3.70\pm0.08)\times 10^{-3}.  \label{eq:CKM:MZ}
\end{align}
Moreover, the Jarlskog invariant~\cite{Jarlskog:1985ht} is reported to be~\cite{UTfit:2022hsi} 
\begin{align}
    J=(3.09\pm 0.07)\times 10^{-5}.
\end{align} 
We use these input values of the quark mixing parameters at the $M_Z$-scale. It is worth pointing out that the previous UTfit results used in the study of Ref.~\cite{Antusch:2013jca}, when compared with the updated values \cite{UTfit:2022hsi}, show that the uncertainties in the quantities $\theta_{12}$, $\theta_{23}$, and $\delta$ have been reduced by factors of 1.5, 1.6, and 2.2, respectively.

The $M_Z$-scale values obtained above will now be used as inputs to compute the running values of the Yukawa couplings and quark mixing parameters at various scales. For this purpose, we consider the following parameterization of the SM Yukawa coupling matrices:
\begin{align}
    Y_u=\textrm{diag}(y_u,y_c,y_t)\,,
    \;\;\;
    Y_d=V_\textrm{CKM}^\dagger(\theta_{12},\theta_{13},\theta_{23},\delta)\,\textrm{diag}(y_d,y_s,y_b)\,,
    \;\;\;
    Y_e=\textrm{diag}(y_e,y_\mu,y_\tau)\,.
\end{align}
To incorporate the two-loop renormalization group equation (RGE) running from $M_Z$ up to higher scales, we use the \texttt{Mathematica} package \texttt{REAP}\ \cite{Antusch:2005gp}. As before, by sampling over 100,000 points and assuming Gaussian errors, the results for the SM running parameters in the $\overline{\text{MS}}$ renormalization scheme are presented in Tables~\ref{MS, 2022} and~\ref{MS, 2024}, corresponding to the 2022 and 2024 PDG data, respectively.  For the convenience of model builders, we provide the running parameters at several benchmark scales: $10^3$, $3 \times 10^3$, $10^4$, $10^5$, $10^7$, $10^9$, $10^{12}$, and $10^{16}$ GeV.

\section{MSSM running parameters}\label{sec:03}
This section is devoted to determining the Yukawa couplings and quark mixing parameters at the GUT-scale (which we fix at 
$M_\mathrm{GUT} = 2 \times 10^{16}$~GeV) within the MSSM framework. To this end, one requires input values of these parameters at the SUSY-scale in the $\overline{\text{DR}}$ renormalization scheme. Therefore, the running values are converted from the $\overline{\text{MS}}$ to the $\overline{\text{DR}}$ scheme~\cite{Martin:1993yx} for selected energy scales of 1, 3, 10, and 100~TeV, and listed in Tables~\ref{DR, 2022} and \ref{DR, 2024} for the 2022 and 2024 PDG data sets, respectively.

In order to compute the parameters at high energies within a SUSY framework, the SM must be matched to its supersymmetric extension. A widely used approximation -- also adopted in this work -- is to perform the matching at a single threshold scale, $M_\mathrm{SUSY}$, typically defined as the geometric mean of two stop masses. At this scale all superpartners are integrated out simultaneously. For moderate or large values of $\tan\beta$, the radiative threshold corrections can become significant, as certain loop diagrams receive $\tan\beta$-enhanced contributions~\cite{Hall:1993gn,Carena:1994bv,Hempfling:1993kv,Blazek:1995nv,Antusch:2008tf,Crivellin:2012zz,Antusch:2015nwi}. The precise value of these threshold corrections depends on the details of the mass spectrum of the superpartners. In this work we give the running parameters at high-scale with zero SUSY loop-threshold corrections, and present an approximate way to take them into account a posteriori.
The suggested procedure for this is discussed towards the end of this section. 

For parameterizing the SUSY threshold corrections, to be applied at the matching scale $M_\mathrm{SUSY}$ we use a similar notation as in~\cite{Antusch:2013jca} 
\begin{align}
    &Y_u^\textrm{SM}=Y_u^\textrm{MSSM}\sin\beta\,, \label{eq:01}\\
    &Y_d^\textrm{SM}=(\mathbb{1}+\text{diag}(\eta_q,\eta_q,\eta_b))Y_d^\textrm{MSSM}\cos\beta\,,\\
    &Y_e^\textrm{SM}=(\mathbb{1}+\text{diag}(\eta_\ell,\eta_\ell,\eta_\tau))Y_e^\textrm{MSSM}\cos\beta\,, \label{eq:02}
\end{align}
where we use $Y_f^\textrm{SM}= Y_f$ to distinguish the SM Yukawa matrices from the MSSM ones. 
Here, we only consider $\tan\beta$ enhanced threshold corrections and assume that the threshold corrections to the first two generations are identical, which holds to a very good approximation in many SUSY models. The threshold corrections for the third family are generically different, since they also include diagrams with 33-elements of the trilinear scalar interactions.\footnote{We note that the parameter $\eta_\tau$ can be absorbed into $\tan\beta$ via $\cos\overline\beta= \cos\beta \; (1+\eta_\tau)$ 
as discussed in Ref.~\cite{Antusch:2013jca}, with the remaining $\eta$’s redefined accordingly. This parameterization reduces the number of independent parameters in the system of Eqs.~\eqref{eq:01}--\eqref{eq:02} by one. In this work, however, we do not employ this possible reparametrization.}

To provide some guidance regarding typical values for the threshold corrections parameters, their origin from SUSY-loop diagrams and example ranges for their typical sizes can be found e.g.\ in \cite{Antusch:2008tf}, where the $\eta$’s are written as $\eta_i = \varepsilon_i \tan\beta$. The $\varepsilon_i$ stem from loop diagrams, and are typically below the percent level. However due to the enhancement by the factor $\tan\beta$, which arises from coupling the down quarks to the Higgs doublet $H_u$ instead of $H_d$, the $\eta_i$ can become as large as ${\pm \,\cal O}(0.5)$ for the example of $\tan\beta = 50$.\footnote{Even larger values for the threshold corrections can be possible for specific choices of MSSM parameters. Also, note that the size of the threshold corrections remains unchanged if all mass-dimensionfull MSSM parameters are scaled to larger values by a common factor. This implies that the example ranges provided in \cite{Antusch:2008tf} also apply to heavier (rescaled) SUSY spectra.} For smaller $\tan \beta$ their typical sizes are reduced accordingly. The leading contributions typically stem from the gluino-loop diagram that affects $\varepsilon_q$ and $\varepsilon_b$, and also to some extent from the diagram proportional to the trilinear soft SUSY breaking term $A_t$,
which affects only $\varepsilon_b$. The threshold corrections in the lepton sector are typically smaller than in the quark sector due to the absence of these two contributions. 

Regarding the RG evolution above the SUSY-scale, we note that since $\eta_q$ and $\eta_\ell$ only imply corrections to Yukawa couplings of the first two families, which are comparatively small, their effects can be neglected in the beta-functions when calculating the RG evolution. This means that to a good approximation, the RG evolution above $M_\mathrm{SUSY}$ only depends on $\eta_b$, $\eta_\tau$, and $\tan \beta$.

For vanishing threshold corrections, the matching conditions between the SM and MSSM Yukawa couplings take the following simple form:
\begin{align}
    Y_u^\textrm{SM}=Y_u^\textrm{MSSM}\sin\beta\,,
    \quad
    Y_d^\textrm{SM}=Y_d^\textrm{MSSM}\cos\beta\,,
    \quad
    Y_e^\textrm{SM}=Y_e^\textrm{MSSM}\cos\beta\,.
\end{align}
Using \texttt{REAP}~\cite{Antusch:2005gp}, which implements the two-loop RGEs, we run the MSSM Yukawa couplings from the low scales $M_\text{SUSY}=3$\ TeV and 10\, TeV to the high scale $M_\text{GUT}$ for various choices of $\tan\beta$. In particular, considering  $\tan\beta= 5, 10, 30, 50$ and assuming Gaussian errors for the input parameters, we obtain the results by sampling over 100,000 points. At the high-scale we list the ``SM quantities'', i.e.\ the quantities $y_{u,c,t}^\text{SM}=y_{u,c,t}^\text{MSSM}\sin\beta$, $y_{d,s,b}^\text{SM}=y_{d,s,b}^\text{MSSM}\cos\beta$, $y_{e,\mu,\tau}^\text{SM}=y_{e,\mu,\tau}^\text{MSSM}\cos\beta$, $\theta_{ij}^\text{SM}=\theta_{ij}^\text{MSSM}$, and $\delta^\text{SM}=\delta^\text{MSSM}$. 
The results are presented in Tables~\ref{2022, GUT, 3TeV} and \ref{2024, GUT, 3TeV} for $M_\text{SUSY} = 3$~TeV using the 2022 and 2024 PDG data sets, respectively. Similarly, Tables~\ref{2022, GUT, 10TeV} and \ref{2024, GUT, 10TeV} show the corresponding results for $M_\text{SUSY} = 10$~TeV.

Finally, we provide a prescription for approximately incorporating threshold corrections into our results at the GUT-scale. Obviously, threshold corrections should, in general, be applied at the scale $M_\mathrm{SUSY}$, prior to running the parameters to the GUT-scale. However, for small values of $\tan\beta$ or sufficiently small threshold correction parameters, it can be a reasonably good approximation to first run the uncorrected parameters to the GUT-scale and then apply the threshold corrections there. We therefore propose (under the above-specified conditions) to directly incorporate the threshold corrections to our results listed in Tables\ \ref{2022, GUT, 3TeV}, \ref{2024, GUT, 3TeV}, \ref{2022, GUT, 10TeV}, and \ref{2024, GUT, 10TeV} as follows: 
\begin{align}
    &y_{u,c,t}^\text{SM,th-corr}=y_{u,c,t}^\text{SM}\,, \label{eq:th:01}
    \\
    &y_{d,s}^\text{SM,th-corr}=\frac{1}{1+\eta_q} y_{d,s}^\text{SM}\,,
    \quad
    y_{b}^\text{SM,th-corr}=\frac{1}{1+\eta_b} y_{b}^\text{SM}\,,
    \\&
    y_{e,\mu}^\text{SM,th-corr}=\frac{1}{1+\eta_\ell} y_{e,\mu}^\text{SM}\,,
    \quad
    y_{\tau}^\text{SM,th-corr}=\frac{1}{1+\eta_\tau} y_{\tau}^\text{SM}\,,
    \\
    &
    \theta_{12}^\text{SM,th-corr}=\theta_{12}^\text{SM}\,,
    \quad
    \theta_{i3}^\text{SM,th-corr}=\frac{1+\eta_b}{1+\eta_q}\theta_{i3}^\text{SM}\,,
    \quad
    \delta^\text{SM,th-corr}= \delta^\text{SM}\,.  \label{eq:mixing}
\end{align}
From our numerical analysis we find that for $\tan\beta = 5$, performing the threshold corrections at the GUT-scale as suggested above, introduces only sub-percent-level errors for the GUT-scale Yukawa couplings. The same holds for $\tan\beta = 10$ with $-0.3 < \eta_b$. For $\tan\beta = 30$, the errors remain below 5\% for $-0.2 < \eta_b < 0.4$. For larger $\tan\beta$, the procedure can only be recommended if $\eta_b$ is small (e.g.\ errors below 5\% for $\tan\beta=50$ requires $-0.04<\eta_b<0.06$), and in general a careful treatment that properly includes the threshold corrections already at $M_\mathrm{SUSY}$ is strongly advised. Note that for larger $\tan\beta$ and large negative $\eta_b$, the Yukawa couplings can even become non-perturbative before the GUT-scale (cf.~Ref.~\cite{Antusch:2013jca})---an effect that is clearly missed in the approximate treatment, indicating that it fails dramatically when approaching this regime.

\section{Implications}\label{sec:04}
In this section, we discuss examples for implications of the updated running fermion masses and mixing parameters for model building. In particular, the improved precision and reduced uncertainties of the 2024 PDG and the 2023 UTfit values lead to significantly sharper constraints on theoretical frameworks that aim to explain the flavour structure of the Standard Model and its extensions. 

\subsection{GUT-scale Yukawa ratios and threshold corrections}
Interesting relations between Yukawa couplings can arise in quark-lepton unifying theories such as Pati-Salam\ \cite{Pati:1973rp,Pati:1974yy} or GUTs\ \cite{Georgi:1974sy,Georgi:1974yf, Georgi:1974my,Fritzsch:1974nn} at the unification-scale. Typical examples in $SU(5)$ models include  the Georgi-Glashow relation $y_\tau/y_b=1$\ \cite{Georgi:1974sy}, the Georgi-Jarlskog relation $y_\mu/y_s=3$\ \cite{Georgi:1979df}, or relations arising from higher dimensional operators such as $y_\tau/y_b=3/2$, $y_\mu/y_s=9/2$, or $y_\mu/y_s=6$\ \cite{Antusch:2009gu,Antusch:2013rxa}. 
Given the updated running Yukawa couplings, it is interesting to examine which threshold corrections are required to render such GUT-scale relations viable.

The GUT-scale values of the ratios $y_e/y_d$ and $y_\mu/y_s$ are, to a very good approximation, independent of $\tan\beta$ and $\eta_b$. We obtain the following results at the GUT-scale $M_\textrm{GUT}=2\times 10^{16}$\ GeV: 
\begin{align}\label{thr1}
    &\frac{(1+\eta_\ell)y_e}{(1+\eta_q)y_d}=
    \begin{cases}
        0.397^{+0.026}_{-0.028}\;,&M_\textrm{SUSY}=3\ \text{TeV},\phantom{10} \text{2022 PDG},\\
        0.395^{+0.026}_{-0.028}\;,&M_\textrm{SUSY}=10\ \text{TeV},\phantom{3} \text{2022 PDG},\\
        0.410\pm 0.007\;,&M_\textrm{SUSY}=3\ \text{TeV},\phantom{10} \text{2024 PDG},\\
        0.408\pm 0.007\;,&M_\textrm{SUSY}=10\ \text{TeV},\phantom{3} \text{2024 PDG},
    \end{cases}
    \\&
    \frac{(1+\eta_\ell)y_\mu}{(1+\eta_q)y_s}=
    \begin{cases}
        4.25^{+0.20}_{-0.34}\;,&\quad M_\textrm{SUSY}=3\ \text{TeV},\phantom{10} \text{2022 PDG},\\
        4.24^{+0.20}_{-0.33}\;,&\quad M_\textrm{SUSY}=10\ \text{TeV},\phantom{3} \text{2022 PDG},\\
        4.33\pm 0.06\;,&\quad M_\textrm{SUSY}=3\ \text{TeV},\phantom{10} \text{2024 PDG},\\
        4.31\pm 0.06\;,&\quad M_\textrm{SUSY}=10\ \text{TeV},\phantom{3} \text{2024 PDG}.
    \end{cases}
\end{align}
Motivated e.g.\ by the findings of Ref.~\cite{Antusch:2025rbp}, which shows that 
\(M_\mathrm{SUSY}\sim\mathcal{O}(5\!-\!10)\,\mathrm{TeV}\) is required to correctly reproduce the 
observed Standard Model Higgs mass with CMSSM boundary conditions, we have chosen 
SUSY-scales of 3 and 10~TeV for the quantities quoted above.

For the third family, the GUT-scale ratio $y_\tau/y_b$ depends on $\tan\beta$, and it is only independent (to a reasonably good approximation) of $\eta_b$ if $\tan\beta$ is not too large and/or if $\eta_b$ is small enough, as discussed at the end of section\ \ref{sec:03} (see also e.g.\ Figure 6(b) in Ref.\ \cite{Antusch:2013jca}). For the respective values $\tan\beta=\lbrace\, 5,\;10,\;30,\;50 \,\rbrace$ we obtain at $M_\textrm{GUT}=2\times 10^{16}$\ GeV  (for \textbf{2022 PDG} input)
\begin{align}\label{thr3}
     &\frac{(1+\eta_\tau)y_\tau}{(1+\eta_b)y_b}=
     \begin{cases}
         1.306^{+0.014}_{-0.018}\,,\quad
        1.311^{+0.014}_{-0.018}\,,\quad
        1.311^{+0.015}_{-0.019}\,,\quad
        1.314^{+0.018}_{-0.024}\,,\quad
        M_\textrm{SUSY}=3\ \text{TeV},
        \\
        1.322^{+0.014}_{-0.018}\,,\quad
        1.327^{+0.014}_{-0.018}\,,\quad
        1.331^{+0.015}_{-0.019}\,,\quad
        1.346^{+0.018}_{-0.023}\,,\quad
        M_\textrm{SUSY}=10\ \text{TeV},
     \end{cases}
\end{align}
and (for \textbf{2024 PDG} input)  
\begin{align}\label{thr4}
     &\frac{(1+\eta_\tau)y_\tau}{(1+\eta_b)y_b}=
     \begin{cases}
        1.309\pm 0.013\,,\;
        1.313\pm 0.013\,,\;
        1.313\pm 0.014\,,\;
        1.318\pm 0.017\;,\;M_\textrm{SUSY}=3\ \text{TeV},\\
        1.325\pm 0.013\,,\;
        1.330\pm 0.013\,,\;
        1.334\pm 0.014\,,\;
        1.350\pm 0.016\,,\;M_\textrm{SUSY}=10\ \text{TeV}.
     \end{cases}
\end{align}
From Eqs.~\eqref{thr1}–\eqref{thr4}, one can deduce the approximate threshold corrections required to achieve a given GUT-scale Yukawa ratio.
Note that, as discussed in section\ \ref{sec:03}, the quark threshold corrections typically dominate over the ones in the charged lepton sector, and the typical size of the threshold corrections is limited. 

We also like to remark that the constraints which a supersymmetric GUT flavour model can impose on the threshold corrections translates into restrictions on the SUSY parameters (e.g.\ on the superpartner particle masses). Combining them with the measurement of the mass of the SM-like Higgs particle can result in predicted ranges for the SUSY particle spectrum (see e.g.\ \cite{Antusch:2015nwi,Antusch:2017ano,Antusch:2019gmc,Antusch:2025rbp}).

\subsection{Threshold corrections and the GUT-scale CKM mixing angles}

As discussed above, for a given GUT model that is predictive regarding the GUT-scale Yukawa relations, Eqs.~\eqref{thr1}--\eqref{thr4} impose constraints on the threshold corrections required to reproduce the observed down-type quark and charged-lepton masses. 
Assuming $\eta_q \gg \eta_\ell$ and $\eta_b \gg \eta_\tau$, such relations can even allow to determine \(\eta_q\) and \(\eta_b\) and consequently, via Eq.~\eqref{eq:mixing}, also the values of the quark mixing angles \(\theta_{13}\) and \(\theta_{23}\) to be fitted at the GUT-scale.  
This can lead to interesting correlations between fermion Yukawa couplings and mixing parameters, and more generally implies that threshold corrections cannot be arbitrarily chosen to fit only the Yukawa couplings.  

Interestingly, because both mixing angles share the same dependence on the threshold corrections, the ratio \(\theta_{13}/\theta_{23}\) is effectively independent of them. Furthermore, it is stable under RG evolution to a very good approximation.  
Hence, using the most recent precise measurements, we obtain:
\begin{align}
\frac{\theta_{13}}{\theta_{23}} = (8.824\pm 0.209) \times 10^{-2} ,
    && 
    \text{(UTfit 2023)}.
\end{align}
This constraint can provide a powerful discriminator between flavour models. It can directly be applied at the high-energy-scale, where the models are defined.

\subsection{The first family Yukawa double ratio -- another powerful model discriminator}
Another great model discriminator is the first family double Yukawa ratio \cite{Antusch:2013jca}. It is highly stable under RG running and threshold corrections, and we obtain 
\begin{align}
    &
    \frac{y_\mu}{y_s}\frac{y_d}{y_e}=
    10.39^{+1.12}_{-0.83}\,,
    && 
    \text{(PDG 2022)},
    \\
    &
    \frac{y_\mu}{y_s}\frac{y_d}{y_e}=
    10.58\pm 0.11\,,
    && 
    \text{(PDG 2024)}.
\end{align}
With the reduced uncertainties of the PDG 2024 data set, this constraint can be particularly challenging to satisfy.

To illustrate this, let us consider the Yukawa submatrices of the first two families in a SU(5) GUT flavour model where the down-quark and charged lepton Yukawa matrices are diagonal and dominated by single higher-dimensional GUT operators (i.e.\ ``single operator dominance''). The entries of the Yukawa submatrices are then related by group theoretical Clebsch-Gordan factors $c_e$ and $c_\mu$, and we can write
\begin{align}
Y_d &= \mathrm{diag}(y_d, y_s)\,, \qquad
Y_e = \mathrm{diag}(c_e\, y_d,\, c_\mu\, y_s)\,.
\end{align}
Under this assumption, the combination of Clebsches that is most favorable regarding the double ratio constraint is \((c_e, c_\mu) = (4/9,\, 9/2)\), as discussed in \cite{Antusch:2013rxa}, which yields
\begin{align}
\frac{y_\mu}{y_s}\frac{y_d}{y_e} = 10.125\,.
\end{align}
While this value remains compatible with the PDG 2022 range, the PDG 2024 result -- with its significantly reduced uncertainty -- places it in tension at more than \(4\sigma\). 
Given the high accuracy of the PDG 2024 result, however, it will be important to carefully check also  corrections to GUT-scale predictions that are induced, e.g., by effective operators at higher order, since they may now dominate over the reduced experimental uncertainties.

\subsection{The GST relation at a new level of accuracy}
Regarding mixing angle predictions, if the Yukawa matrix $Y_d$ is hierarchical and has a symmetric 1-2 block with a vanishing 1-1 element, one can derive the approximate Gatto-Sartori-Tonin (GST) relation $\sqrt{y_d/y_s}\approx \theta_{12}$\ \cite{Gatto:1968ss}, when further assuming that the Cabibbo angle is dominated by the down quark sector. By an analytical diagonalization we find the exact relation $\arctan(\sqrt{y_d/y_s})= \theta_{12}$.\footnote{Predictions for the Yukawa ratio $\sqrt{y_d/y_s}$ also emerge under other assumptions, e.g.\ in the $SU(5)$ GUT scenario discussed in \cite{Antusch:2018gnu}.} 
For this quantity, which is again highly stable under RG running and SUSY threshold corrections, we obtain 
\begin{align}
    &
    \arctan\left(\sqrt{\frac{y_d}{y_s}}\right)
    =0.2190^{+0.0112}_{-0.0085}\,,
    &&
    \text{(PDG 2022)},
    \\
    &
    \arctan\left(\sqrt{\frac{y_d}{y_s}}\right)
    =0.2205\pm 0.0012\,,
    &&
    \text{(PDG 2024)}.
\end{align}
The experimental central value for the Cabibbo angle $\theta_{12}$ given in Eq.\ \eqref{eq:CKM:MZ} lies well within the 1$\sigma$ range of the PDG 2022 result. However, with the 2024 data, the uncertainty intervals only overlap at the 3$\sigma$ level.
We also note that with the reduced uncertainties of the PDG 2024 dataset, the approximate GST relation is not accurate enough. Furthermore, we would like to remark that when confronting the PDG 2024 results with models, it is important to carefully account for uncertainties arising from the model construction and analysis. 

\subsection{Quark mixing and phase sum rules}

Motivated by the smallness of \(\theta_{13}\) in CKM matrix, one may postulate that the 1-3 mixing could indeed be zero in both the up-quark and the down-quark sectors (cf.\ \cite{Antusch:2022ufb}).  
The small 1-3 mixing in the CKM matrix is then induced by the non-commutativity of the 1-2 and 2-3 rotations for diagonalizing the up-quark and down-quark Yukawa matrices. An approximate realization of this scenario can be achieved by texture zeros in the 1-3 positions the of Yukawa matrices. 

With this hypothesis, one can get, for instance, a guideline for how large the 1-2 mixings in the up- and down-quark sectors typically are, and potentially some insight about the possible origin of CP violation in the quark Yukawa sector. Starting with the 1-2 mixings, the following exact ``quark mixing sum rules'' hold in the up- and down-quark left-mixing matrices in the flavour basis\ \cite{Antusch:2022ufb,Antusch:2009hq} (see also\ \cite{Ballett:2014dua,Marzocca:2013cr,Petcov:2014laa}):
\begin{align}\label{eq:mixing sum rule}
    t_{12}^u=\frac{t_{13}}{s_{23}}\,,
    \qquad
    t_{12}^d=
   \left|\frac{t_{12}t_{23}-s_{13}e^{i\delta}}{t_{23}+t_{12}s_{13}e^{i\delta}}\right|,
\end{align}
where we follow the notation in Ref.\ \cite{Antusch:2022ufb}. The 1-2 up-left-mixing angle $\theta_{12}^u$ that appears in the first relation in Eq.\ \eqref{eq:mixing sum rule} is approximately invariant under RG running as well as under threshold corrections. By utilizing the UTfit 2023 data, we obtain
\begin{align}
    t_{12}^u=0.08801\pm0.00209\,,\qquad
    \theta_{12}^u=5.043^\circ\pm0.124^\circ .
\end{align}
The 1-2 down-left-mixing angle $\theta_{12}^d$, which appears in the second relation in Eq.\ \eqref{eq:mixing sum rule}, depends on RG running, but is independent of $\tan\beta$. We find 
\begin{align}
   &t_{12}^d=0.2063\pm0.0022\,,
   &&
    \theta_{12}^d=11.66^\circ\pm0.12^\circ
    &&
    \text{at $M_Z$},
    \\
    &t_{12}^d=0.2082\pm0.0022\,,
    &&
    \theta_{12}^d=11.76^\circ\pm0.12^\circ
    &&
    \text{at $M_\textrm{GUT}$},
\end{align}
where the GUT-scale value holds for RG running in the SM as well as in the MSSM (independent of $\tan\beta$).

Moreover, if additionally all up-quark left-mixing angles are small, the following  ``quark phase sum rule'' can be derived\ \cite{Antusch:2022ufb,Antusch:2009hq}: 
\begin{align}\label{eq:phase sum rule}
    \delta_{12}^d-\delta_{12}^u=\arg\left(
    \frac{s_{13}-t_{12}t_{23}e^{-i\delta}}{t_{23}+t_{12}s_{13}e^{i\delta}}    
    \right).
\end{align}
It allows to relate the observable CKM Dirac CP phase $\delta$ to the phase difference of the up- and down-quark 1-2 left-rotations. The phase difference $\delta_{12}^d-\delta_{12}^u$ required to match the observed result for $\delta$ depends somewhat on RG running, and we obtain 
\begin{align}
   &\delta_{12}^d-\delta_{12}^u=91.06^\circ\pm1.48^\circ
   \qquad\text{at $M_Z$},
   \\
    &\delta_{12}^d-\delta_{12}^u=91.29^\circ\pm1.48^\circ
   \qquad\text{at $M_\textrm{GUT}$}.
\end{align}
Again, the GUT-scale result holds for RG running in both the SM as well as the MSSM (independent of $\tan\beta$).
Interestingly, with the updated more precise UTfit results, a phase difference of $90^\circ$ lies well within the 1$\sigma$ region.\footnote{We remark that there is an approximate relation between this phase difference and the UT angle $\alpha$ (cf. \cite{Antusch:2009hq}), which is also close to $90^\circ$. However, this relation is not accurate enough with respect to the updated UTfit results. While the UTfit 2023 results\ \cite{UTfit:2022hsi} correspond to $\alpha = 92.4^\circ\pm 1.4^\circ$, where $90^\circ$ is outside the $1\sigma$ range, the exact ``phase sum rule'' with $90^\circ$ leads to the correct amount of CP violation at this confidence level.} 
Such a phase difference of $90^\circ$ might originate from a single imaginary Yukawa matrix element (cf.\ \cite{Antusch:2022ufb,Antusch:2009hq}), that could emerge, e.g., in models with spontaneous CP violation (see e.g.\ \cite{Antusch:2011sx}).

\subsection{Remark on the role of theoretical uncertainties}
In summary, the above discussion examined the implications of the updated running parameters for Yukawa relations, as well as for predictive frameworks of quark mixing angles and CP violation. In particular, we have shown that, when using the PDG 2024 dataset, these relations can act as powerful discriminators among flavour models. We  provided updated results that allow several of these relations to be applied directly at the GUT scale. It is important to emphasize that, beyond the purely parametric uncertainties -- arising directly from uncertainties in input values such as the PDG parameters -- additional theoretical uncertainties exist that are are inherently model-dependent. Examples are corrections at higher loop orders, effects from unconsidered higher-dimensional operators, or from a more refined treatment of threshold corrections. Since the size of such uncertainties depends on the details of the underlying model, and the analysis of specific flavour models lies outside the scope of this study, we cannot provide estimates for them here. In general, however, the theoretical uncertainties are, at least in some cases, expected to be comparable to, or even larger than, the experimental ones -- especially w.r.t.\ the PDG 2024 dataset. When this is the case, it will be particularly important for model builders to identify which part of the uncertainties can be reduced by an improvement of the model analysis, and which parts are an irreducible consequence of the model construction, finally limiting the possibility to distinguish the prediction of a given model from others.

\section{Conclusions}\label{sec:con}
In this work, we have revisited the running quark and lepton Yukawa couplings, as well as the quark mixing parameters, in light of the most recent Particle Data Group (PDG) releases, namely the 2022 and 2024 data sets. The 2024 PDG determinations of low-energy fermion masses exhibit significantly smaller uncertainties, due to reduced estimates of systematic errors compared to the more conservative 2022 analysis. To evaluate the impact of these changes, we have presented running parameters derived from both the 2022 and 2024 datasets within the frameworks of the Standard Model (SM) and its minimal supersymmetric extension (MSSM). 

Within the framework of the SM, we have presented the running parameters at multiple scales, ranging from $M_Z$ up to the GUT-scale. In the MSSM, we have provided the corresponding parameters at the GUT-scale for various values of $\tan\beta$, considering 3 TeV and 10 TeV as supersymmetry-breaking scales. These updated running parameters, with their refined error estimates, constitute a useful input for model-building efforts, offering more precise constraints for constructing and testing theories beyond the SM. 

We have also discussed implications of the updated running parameters for Yukawa relations, as well as for predictive schemes for quark mixing angles and CP violation. In particular, we have shown that using the PDG 2024 dataset, these relations can serve as powerful discriminators between flavour models, and we have provided updated results that allow some of them to be applied directly at the GUT-scale.  
Our examples demonstrate that some of the simplest flavour relations appear in tension with the PDG 2024 data.  
However, we emphasize the importance of carefully accounting for theoretical uncertainties from next-to-leading order effects in the model construction and analysis, as they may now introduce uncertainties larger than the experimental ones.

\section*{Acknowledgments}
S.S.\ acknowledges the financial support from the Slovenian Research Agency (research core funding No.\ P1-0035 and N1-0321). 

\begin{table}[]
\centering
\renewcommand{\arraystretch}{1.25}
\resizebox{0.95\textwidth}{!}{
 \begin{tabular}{|c|ll|ll|ll|ll|}
\hline
SM Quantity & \multicolumn{2}{c|}{$\mu=1\,$TeV}& \multicolumn{2}{c|}{$\mu=3\,$TeV} & \multicolumn{2}{c|}{$\mu=10\,$TeV} & \multicolumn{2}{c|}{$\mu=100\,$TeV}\\
\hline$y_u/10^{-6}$ & $6.21$ & $^{+1.36}_{-0.78}$ & $5.90$ & $^{+1.30}_{-0.73}$ & $5.60$ & $^{+1.23}_{-0.70}$ & $5.13$ & $^{+1.13}_{-0.64}$  \\ $y_d/10^{-5}$ & $1.36$ & $^{+0.12}_{-0.07}$ & $1.29$ & $^{+0.12}_{-0.06}$ & $1.23$ & $^{+0.11}_{-0.06}$ & $1.12$ & $^{+0.11}_{-0.05}$  \\ $y_s/10^{-4}$ & $2.72$ & $^{+0.22}_{-0.13}$ & $2.58$ & $^{+0.21}_{-0.12}$ & $2.46$ & $^{+0.20}_{-0.12}$ & $2.25$ & $^{+0.18}_{-0.11}$  \\ $y_c/10^{-3}$ & $3.10$ & $^{+0.09}_{-0.08}$ & $2.95$ & $^{+0.08}_{-0.08}$ & $2.80$ & $^{+0.08}_{-0.08}$ & $2.56$ & $^{+0.07}_{-0.07}$  \\ $y_b/10^{-2}$ & $1.402$ & $^{+0.014}_{-0.012}$ & $1.323$ & $^{+0.013}_{-0.012}$ & $1.248$ & $^{+0.012}_{-0.012}$ & $1.131$ & $^{+0.012}_{-0.011}$  \\ $y_t$ & $0.8625$ & $^{+0.0043}_{-0.0043}$ & $0.8250$ & $^{+0.0044}_{-0.0043}$ & $0.7890$ & $^{+0.0045}_{-0.0044}$ & $0.7320$ & $^{+0.0046}_{-0.0045}$  \\ \hline
$\theta_{12}$ & $0.22704$ & $^{+0.00083}_{-0.00082}$ & $0.22705$ & $^{+0.00083}_{-0.00082}$ & $0.22705$ & $^{+0.00083}_{-0.00082}$ & $0.22705$ & $^{+0.00083}_{-0.00081}$  \\ $\theta_{23}/10^{-2}$ & $4.275$ & $^{+0.042}_{-0.042}$ & $4.307$ & $^{+0.043}_{-0.041}$ & $4.339$ & $^{+0.042}_{-0.043}$ & $4.394$ & $^{+0.043}_{-0.043}$  \\ $\theta_{13}/10^{-3}$ & $3.77$ & $^{+0.08}_{-0.08}$ & $3.80$ & $^{+0.08}_{-0.08}$ & $3.83$ & $^{+0.08}_{-0.08}$ & $3.88$ & $^{+0.08}_{-0.09}$  \\ $\delta$ & $1.139$ & $^{+0.023}_{-0.023}$ & $1.139$ & $^{+0.023}_{-0.023}$ & $1.139$ & $^{+0.023}_{-0.023}$ & $1.139$ & $^{+0.023}_{-0.023}$  \\
$J/10^{-5}$ & 3.09 & $^{+0.08}_{-0.08}$ & 3.28 & $^{+0.08}_{-0.08}$ & 3.33 & $^{+0.09}_{-0.09}$ & 3.41 & $^{+0.09}_{-0.09}$ \\
\hline
$y_e/10^{-6}$ & $2.8229$ & $^{+0.0006}_{-0.0006}$ & $2.8405$ & $^{+0.0014}_{-0.0014}$ & $2.8495$ & $^{+0.0015}_{-0.0015}$ & $2.8641$ & $^{+0.0023}_{-0.0024}$  \\ $y_{\mu}/10^{-4}$ & $5.9467$ & $^{+0.0014}_{-0.0012}$ &     $5.9839$ & $^{+0.0030}_{-0.0029}$ & $6.0028$ & $^{+0.0032}_{-0.0032}$ & $6.0337$ & $^{+0.0049}_{-0.0051}$  \\ $y_{\tau}/10^{-2}$ & $1.0101$ & $^{+0.0003}_{-0.0002}$ &  $1.0164$ & $^{+0.0005}_{-0.0005}$ & $1.0195$ & $^{+0.0006}_{-0.0005}$ & $1.0248$ & $^{+0.0010}_{-0.0007}$  \\ \hline
$g_3$ & $1.0581$ & $^{+0.0030}_{-0.0031}$ & $1.0046$ & $^{+0.0026}_{-0.0027}$ & $0.9545$ & $^{+0.0022}_{-0.0023}$ & $0.8765$ & $^{+0.0017}_{-0.0018}$  \\ $g_2$ & $0.63811$ & $^{+0.00004}_{-0.00003}$ & $0.63246$ & $^{+0.00004}_{-0.00003}$ & $0.62644$ & $^{+0.00003}_{-0.00003}$ & $0.61539$ & $^{+0.00003}_{-0.00003}$  \\ $g_1$ & $0.467451$ & $^{+0.000026}_{-0.000028}$ & $0.470392$ & $^{+0.000028}_{-0.000027}$ & $0.473680$ & $^{+0.000027}_{-0.000029}$ & $0.480165$ & $^{+0.000030}_{-0.000029}$  \\ \hline
\end{tabular}
}

\vspace{5mm}

\resizebox{0.95\textwidth}{!}{
\begin{tabular}{|c|ll|ll|ll|ll|}
\hline
SM Quantity & \multicolumn{2}{c|}{$\mu=10^7\,$GeV}& \multicolumn{2}{c|}{$\mu=10^9\,$GeV} & \multicolumn{2}{c|}{$\mu=10^{12}\;$GeV} & \multicolumn{2}{c|}{$\mu=10^{16}\;$GeV}\\
\hline$y_u/10^{-6}$ & $4.44$ & $^{+0.98}_{-0.55}$ & $3.95$ & $^{+0.86}_{-0.50}$ & $3.41$ & $^{+0.73}_{-0.44}$ & $2.89$ & $^{+0.63}_{-0.37}$  \\ $y_d/10^{-5}$ & $0.97$ & $^{+0.09}_{-0.04}$ & $0.87$ & $^{+0.08}_{-0.04}$ & $0.76$ & $^{+0.07}_{-0.03}$ & $0.65$ & $^{+0.06}_{-0.03}$  \\ $y_s/10^{-4}$ & $1.96$ & $^{+0.16}_{-0.10}$ & $1.75$ & $^{+0.14}_{-0.09}$ & $1.52$ & $^{+0.12}_{-0.08}$ & $1.31$ & $^{+0.10}_{-0.07}$  \\ $y_c/10^{-3}$ & $2.22$ & $^{+0.06}_{-0.06}$ & $1.97$ & $^{+0.06}_{-0.06}$ & $1.70$ & $^{+0.05}_{-0.05}$ & $1.44$ & $^{+0.05}_{-0.04}$  \\ $y_b/10^{-2}$ & $0.963$ & $^{+0.010}_{-0.010}$ & $0.845$ & $^{+0.010}_{-0.009}$ & $0.720$ & $^{+0.009}_{-0.007}$ & $0.607$ & $^{+0.008}_{-0.006}$  \\ $y_t$ & $0.6469$ & $^{+0.0046}_{-0.0049}$ & $0.5844$ & $^{+0.0049}_{-0.0048}$ & $0.5144$ & $^{+0.0051}_{-0.0047}$ & $0.4462$ & $^{+0.0049}_{-0.0048}$  \\ \hline
$\theta_{12}$ & $0.22706$ & $^{+0.00083}_{-0.00081}$ & $0.22706$ & $^{+0.00083}_{-0.00081}$ & $0.22707$ & $^{+0.00082}_{-0.00082}$ & $0.22708$ & $^{+0.00082}_{-0.00082}$  \\ $\theta_{23}/10^{-2}$ & $4.486$ & $^{+0.044}_{-0.044}$ & $4.561$ & $^{+0.044}_{-0.045}$ & $4.651$ & $^{+0.046}_{-0.046}$ & $4.745$ & $^{+0.046}_{-0.048}$  \\ $\theta_{13}/10^{-3}$ & $3.96$ & $^{+0.08}_{-0.09}$ & $4.03$ & $^{+0.08}_{-0.09}$ & $4.11$ & $^{+0.09}_{-0.09}$ & $4.19$ & $^{+0.09}_{-0.09}$  \\ $\delta$ & $1.139$ & $^{+0.023}_{-0.023}$ & $1.139$ & $^{+0.023}_{-0.023}$ & $1.139$ & $^{+0.023}_{-0.023}$ & $1.139$ & $^{+0.023}_{-0.023}$  \\
$J/10^{-5}$ & 3.56 & $^{+0.09}_{-0.09}$ & 3.68 & $^{+0.10}_{-0.10}$ & 3.82 & $^{+0.10}_{-0.10}$ & 3.97 & $^{+0.10}_{-0.10}$ \\ \hline
$y_e/10^{-6}$ & $2.8691$ & $^{+0.0041}_{-0.0039}$ & $2.8520$ & $^{+0.0056}_{-0.0052}$ & $2.7996$ & $^{+0.0072}_{-0.0073}$ & $2.6952$ & $^{+0.0090}_{-0.0093}$  \\ $y_{\mu}/10^{-4}$ & $6.0442$ & $^{+0.0086}_{-0.0082}$ & $6.0082$ & $^{+0.0118}_{-0.0109}$ & $5.8977$ & $^{+0.0152}_{-0.0153}$ & $5.6779$ & $^{+0.0190}_{-0.0196}$  \\ $y_{\tau}/10^{-2}$ & $1.0267$ & $^{+0.0014}_{-0.0015}$ & $1.0206$ & $^{+0.0020}_{-0.0019}$ & $1.0017$ & $^{+0.0026}_{-0.0026}$ & $0.9644$ & $^{+0.0032}_{-0.0033}$  \\ \hline
$g_3$ & $0.7648$ & $^{+0.0011}_{-0.0012}$ & $0.6871$ & $^{+0.0008}_{-0.0009}$ & $0.6052$ & $^{+0.0006}_{-0.0006}$ & $0.5310$ & $^{+0.0004}_{-0.0004}$  \\ $g_2$ & $0.59493$ & $^{+0.00003}_{-0.00003}$ & $0.57639$ & $^{+0.00003}_{-0.00002}$ & $0.55157$ & $^{+0.00003}_{-0.00002}$ & $0.52296$ & $^{+0.00002}_{-0.00001}$  \\ $g_1$ & $0.493975$ & $^{+0.000031}_{-0.000033}$ & $0.509050$ & $^{+0.000034}_{-0.000036}$ & $0.534497$ & $^{+0.000039}_{-0.000042}$ & $0.575238$ & $^{+0.000048}_{-0.000052}$  \\ \hline
\end{tabular}
}
\caption{Running SM parameters and 1$\sigma$ HPD intervals in the $\overline{\text{MS}}$ renormalization scheme using the \textbf{2022 PDG} data as input. }\label{MS, 2022}
\end{table}

\begin{table}[]
\centering
\renewcommand{\arraystretch}{1.4}
\resizebox{0.95\textwidth}{!}{
 \begin{tabular}{|c|ll|ll|ll|ll|}
\hline
SM Quantity & \multicolumn{2}{c|}{$\mu=1\,$TeV}& \multicolumn{2}{c|}{$\mu=3\,$TeV} & \multicolumn{2}{c|}{$\mu=10\,$TeV} & \multicolumn{2}{c|}{$\mu=100\,$TeV}\\
\hline$y_u/10^{-6}$ & $6.15$ & $\pm 0.14$& $5.84$ & $\pm 0.13$& $5.54$ & $\pm 0.13$& $5.07$ & $\pm 0.12$ \\ $y_d/10^{-5}$ & $1.35$ & $\pm 0.02$& $1.28$ & $\pm 0.02$& $1.22$ & $\pm 0.02$& $1.12$ & $\pm 0.02$ \\ $y_s/10^{-4}$ & $2.68$ & $\pm 0.03$& $2.54$ & $\pm 0.03$& $2.42$ & $\pm 0.03$& $2.22$ & $\pm 0.03$ \\ $y_c/10^{-3}$ & $3.11$ & $\pm 0.05$& $2.95$ & $\pm 0.05$& $2.80$ & $\pm 0.05$& $2.57$ & $\pm 0.05$ \\ $y_b/10^{-2}$ & $1.401$ & $\pm 0.009$& $1.321$ & $\pm 0.009$& $1.246$ & $\pm 0.009$& $1.130$ & $\pm 0.008$ \\ $y_t$ & $0.8616$ & $\pm 0.0043$& $0.8242$ & $\pm 0.0043$& $0.7880$ & $\pm 0.0044$& $0.7309$ & $\pm 0.0046$ \\ \hline
$\theta_{12}$ & $0.22704$ & $\pm 0.00082$& $0.22705$ & $\pm 0.00082$& $0.22705$ & $\pm 0.00082$& $0.22705$ & $\pm 0.00082$ \\ $\theta_{23}/10^{-2}$ & $4.275$ & $\pm 0.042$& $4.307$ & $\pm 0.042$& $4.339$ & $\pm 0.042$& $4.394$ & $\pm 0.043$ \\ $\theta_{13}/10^{-3}$ & $3.77$ & $\pm 0.08$& $3.80$ & $\pm 0.08$& $3.83$ & $\pm 0.08$& $3.88$ & $\pm 0.08$ \\ $\delta$ & $1.139$ & $\pm 0.023$& $1.139$ & $\pm 0.023$& $1.139$ & $\pm 0.023$& $1.139$ & $\pm 0.023$ \\ 
$J/10^{-5}$ & 3.23 & $\pm0.08$ & 3.28 & $\pm0.08$ & 3.33 & $\pm0.09$ & 3.41 & $\pm0.09$
\\\hline
$y_e/10^{-6}$ & $2.8227$ & $\pm 0.0006$&  $2.8402$ & $\pm 0.0014$& $2.8492$ & $\pm 0.0015$& $2.8637$ & $\pm 0.0024$ \\ $y_{\mu}/10^{-4}$ & $5.9465$ & $\pm 0.0013$&    $5.9833$ & $\pm 0.0030$& $6.0022$ & $\pm 0.0032$& $6.0327$ & $\pm 0.0050$ \\ $y_{\tau}/10^{-2}$ & $1.0101$ & $\pm 0.0002$&  $1.0163$ & $\pm 0.0005$& $1.0196$ & $\pm 0.0005$& $1.0248$ & $\pm 0.0008$ \\ \hline
$g_3$ & $1.0583$ & $\pm 0.0030$& $1.0049$ & $\pm 0.0026$& $0.9547$ & $\pm 0.0022$& $0.8767$ & $\pm 0.0017$ \\ $g_2$ & $0.63811$ & $\pm 0.00003$& $0.63246$ & $\pm 0.00003$& $0.62644$ & $\pm 0.00003$& $0.61539$ & $\pm 0.00003$ \\ $g_1$ & $0.467453$ & $\pm 0.000027$& $0.470394$ & $\pm 0.000027$& $0.473681$ & $\pm 0.000028$& $0.480166$ & $\pm 0.000029$ \\ \hline
\end{tabular}
}

\vspace{5mm}

\resizebox{0.95\textwidth}{!}{
\begin{tabular}{|c|ll|ll|ll|ll|}
\hline
SM Quantity & \multicolumn{2}{c|}{$\mu=10^7\,$GeV}& \multicolumn{2}{c|}{$\mu=10^9\,$GeV} & \multicolumn{2}{c|}{$\mu=10^{12}\;$GeV} & \multicolumn{2}{c|}{$\mu=10^{16}\;$GeV}\\
\hline$y_u/10^{-6}$ & $4.39$ & $\pm 0.10$& $3.90$ & $\pm 0.09$& $3.37$ & $\pm 0.08$& $2.87$ & $\pm 0.07$ \\ $y_d/10^{-5}$ & $0.97$ & $\pm 0.01$& $0.87$ & $\pm 0.01$& $0.75$ & $\pm 0.01$& $0.65$ & $\pm 0.01$ \\ $y_s/10^{-4}$ & $1.93$ & $\pm 0.03$& $1.72$ & $\pm 0.02$& $1.50$ & $\pm 0.02$& $1.29$ & $\pm 0.02$ \\ $y_c/10^{-3}$ & $2.22$ & $\pm 0.04$& $1.98$ & $\pm 0.04$& $1.71$ & $\pm 0.03$& $1.45$ & $\pm 0.03$ \\ $y_b/10^{-2}$ & $0.961$ & $\pm 0.008$& $0.844$ & $\pm 0.007$& $0.719$ & $\pm 0.006$& $0.606$ & $\pm 0.006$ \\ $y_t$ & $0.6462$ & $\pm 0.0047$& $0.5838$ & $\pm 0.0048$& $0.5138$ & $\pm 0.0048$& $0.4454$ & $\pm 0.0048$ \\ \hline
$\theta_{12}$ & $0.22706$ & $\pm 0.00082$& $0.22706$ & $\pm 0.00082$& $0.22707$ & $\pm 0.00082$& $0.22708$ & $\pm 0.00082$ \\ $\theta_{23}/10^{-2}$ & $4.486$ & $\pm 0.044$& $4.560$ & $\pm 0.045$& $4.650$ & $\pm 0.046$& $4.743$ & $\pm 0.047$ \\ $\theta_{13}/10^{-3}$ & $3.96$ & $\pm 0.09$& $4.03$ & $\pm 0.09$& $4.11$ & $\pm 0.09$& $4.19$ & $\pm 0.09$ \\ $\delta$ & $1.139$ & $\pm 0.023$& $1.139$ & $\pm 0.023$& $1.139$ & $\pm 0.023$& $1.139$ & $\pm 0.023$  \\ 
$J/10^{-5}$ & 3.56 & $\pm0.09$ & 3.68 & $\pm0.10$ & 3.83 & $\pm0.10$ & 3.98 & $\pm0.10$
\\ \hline
$y_e/10^{-6}$ & $2.8681$ & $\pm 0.0039$& $2.8508$ & $\pm 0.0054$& $2.7978$ & $\pm 0.0072$& $2.6935$ & $\pm 0.0091$ \\ $y_{\mu}/10^{-4}$ & $6.0421$ & $\pm 0.0083$& $6.0056$ & $\pm 0.0113$& $5.8939$ & $\pm 0.0152$& $5.6745$ & $\pm 0.0192$ \\ $y_{\tau}/10^{-2}$ & $1.0264$ & $\pm 0.0014$& $1.0201$ & $\pm 0.0019$& $1.0012$ & $\pm 0.0026$& $0.9639$ & $\pm 0.0033$ \\ \hline
$g_3$ & $0.7649$ & $\pm 0.0011$& $0.6872$ & $\pm 0.0008$& $0.6052$ & $\pm 0.0006$& $0.5310$ & $\pm 0.0004$ \\ $g_2$ & $0.59493$ & $\pm 0.00003$& $0.57639$ & $\pm 0.00002$& $0.55157$ & $\pm 0.00002$& $0.52296$ & $\pm 0.00002$ \\ $g_1$ & $0.493977$ & $\pm 0.000032$& $0.509051$ & $\pm 0.000035$& $0.534498$ & $\pm 0.000040$& $0.575240$ & $\pm 0.000050$ \\ \hline
\end{tabular}
}
\caption{Running SM parameters and 1$\sigma$ HPD intervals in the $\overline{\text{MS}}$ renormalization scheme using the \textbf{2024 PDG} data as input. }\label{MS, 2024}
\end{table}
\clearpage

\begin{table}[]
\centering
\renewcommand{\arraystretch}{1.25}
\resizebox{0.95\textwidth}{!}{
 \begin{tabular}{|c|ll|ll|ll|ll|}
\hline
SM Quantity & \multicolumn{2}{c|}{$\mu=1\,$TeV}& \multicolumn{2}{c|}{$\mu=3\,$TeV} & \multicolumn{2}{c|}{$\mu=10\,$TeV} & \multicolumn{2}{c|}{$\mu=100\,$TeV}\\
\hline$y_u/10^{-6}$ & $6.09$ & $^{+1.34}_{-0.76}$ & $5.84$ & $^{+1.28}_{-0.73}$ & $5.51$ & $^{+1.22}_{-0.68}$ & $5.06$ & $^{+1.12}_{-0.63}$  \\ $y_d/10^{-5}$ & $1.33$ & $^{+0.12}_{-0.06}$ & $1.28$ & $^{+0.12}_{-0.06}$ & $1.20$ & $^{+0.11}_{-0.05}$ & $1.10$ & $^{+0.11}_{-0.05}$  \\ $y_s/10^{-4}$ & $2.66$ & $^{+0.22}_{-0.13}$ & $2.56$ & $^{+0.21}_{-0.12}$ & $2.42$ & $^{+0.19}_{-0.12}$ & $2.22$ & $^{+0.18}_{-0.11}$  \\ $y_c/10^{-3}$ & $3.04$ & $^{+0.08}_{-0.08}$ & $2.92$ & $^{+0.08}_{-0.08}$ & $2.75$ & $^{+0.08}_{-0.08}$ & $2.53$ & $^{+0.07}_{-0.07}$  \\ $y_b/10^{-2}$ & $1.373$ & $^{+0.014}_{-0.012}$ & $1.310$ & $^{+0.013}_{-0.012}$ & $1.227$ & $^{+0.012}_{-0.012}$ & $1.115$ & $^{+0.012}_{-0.011}$  \\ $y_t$ & $0.8472$ & $^{+0.0041}_{-0.0044}$ & $0.8171$ & $^{+0.0044}_{-0.0043}$ & $0.7775$ & $^{+0.0046}_{-0.0043}$ & $0.7230$ & $^{+0.0046}_{-0.0045}$  \\ \hline
$\theta_{12}$ & $0.22704$ & $^{+0.00083}_{-0.00082}$ & $0.22705$ & $^{+0.00083}_{-0.00082}$ & $0.22705$ & $^{+0.00083}_{-0.00082}$ & $0.22705$ & $^{+0.00083}_{-0.00081}$  \\ $\theta_{23}/10^{-2}$ & $4.275$ & $^{+0.042}_{-0.042}$ & $4.307$ & $^{+0.043}_{-0.041}$ & $4.339$ & $^{+0.042}_{-0.043}$ & $4.394$ & $^{+0.043}_{-0.043}$  \\ $\theta_{13}/10^{-3}$ & $3.77$ & $^{+0.08}_{-0.08}$ & $3.80$ & $^{+0.08}_{-0.08}$ & $3.83$ & $^{+0.08}_{-0.08}$ & $3.88$ & $^{+0.08}_{-0.09}$  \\ $\delta$ & $1.139$ & $^{+0.023}_{-0.023}$ & $1.139$ & $^{+0.023}_{-0.023}$ & $1.139$ & $^{+0.023}_{-0.023}$ & $1.139$ & $^{+0.023}_{-0.023}$   \\ 
$J/10^{-5}$ & 3.09 & $^{+0.08}_{-0.08}$ & 3.28 & $^{+0.08}_{-0.08}$ & 3.33 & $^{+0.09}_{-0.09}$ & 3.41 & $^{+0.09}_{-0.09}$ 
\\ \hline
$y_e/10^{-6}$ & $2.8280$ & $^{+0.0007}_{-0.0007}$ & $2.8397$ & $^{+0.0011}_{-0.0011}$ & $2.8531$ & $^{+0.0016}_{-0.0016}$ & $2.8667$ & $^{+0.0024}_{-0.0025}$  \\ $y_{\mu}/10^{-4}$ & $5.9576$ & $^{+0.0015}_{-0.0014}$ & $5.9822$ & $^{+0.0023}_{-0.0022}$ & $6.0105$ & $^{+0.0033}_{-0.0035}$ & $6.0391$ & $^{+0.0051}_{-0.0053}$  \\ $y_{\tau}/10^{-2}$ & $1.0119$ & $^{+0.0003}_{-0.0002}$ & $1.0161$ & $^{+0.0004}_{-0.0003}$ & $1.0208$ & $^{+0.0007}_{-0.0005}$ & $1.0257$ & $^{+0.0010}_{-0.0008}$  \\ \hline
$g_3$ & $1.0486$ & $^{+0.0029}_{-0.0030}$ & $1.0044$ & $^{+0.0025}_{-0.0027}$ & $0.9475$ & $^{+0.0021}_{-0.0023}$ & $0.8711$ & $^{+0.0017}_{-0.0017}$  \\ $g_2$ & $0.63746$ & $^{+0.00004}_{-0.00003}$ & $0.63273$ & $^{+0.00004}_{-0.00003}$ & $0.62582$ & $^{+0.00003}_{-0.00003}$ & $0.61480$ & $^{+0.00003}_{-0.00003}$  \\ $g_1$ & $0.468063$ & $^{+0.000026}_{-0.000028}$ & $0.470535$ & $^{+0.000028}_{-0.000027}$ & $0.474317$ & $^{+0.000029}_{-0.000028}$ & $0.480828$ & $^{+0.000030}_{-0.000029}$  \\ \hline
\end{tabular}
}
\caption{Running SM parameters and 1$\sigma$ HPD intervals in the $\overline{\text{DR}}$ renormalization scheme using the \textbf{2022 PDG} data as input. }\label{DR, 2022}

\vspace{5mm}

\renewcommand{\arraystretch}{1.4}
\resizebox{0.95\textwidth}{!}{
 \begin{tabular}{|c|ll|ll|ll|ll|}
\hline
SM Quantity & \multicolumn{2}{c|}{$\mu=1\,$TeV}& \multicolumn{2}{c|}{$\mu=3\,$TeV} & \multicolumn{2}{c|}{$\mu=10\,$TeV} & \multicolumn{2}{c|}{$\mu=100\,$TeV}\\
\hline$y_u/10^{-6}$ & $6.03$ & $\pm 0.13$& $5.78$ & $\pm 0.13$& $5.45$ & $\pm 0.12$& $5.01$ & $\pm 0.11$ \\ $y_d/10^{-5}$ & $1.32$ & $\pm 0.02$& $1.27$ & $\pm 0.02$& $1.20$ & $\pm 0.02$& $1.10$ & $\pm 0.02$ \\ $y_s/10^{-4}$ & $2.63$ & $\pm 0.03$& $2.52$ & $\pm 0.03$& $2.38$ & $\pm 0.03$& $2.19$ & $\pm 0.03$ \\ $y_c/10^{-3}$ & $3.05$ & $\pm 0.05$& $2.92$ & $\pm 0.05$& $2.76$ & $\pm 0.05$& $2.53$ & $\pm 0.05$ \\ $y_b/10^{-2}$ & $1.372$ & $\pm 0.009$& $1.308$ & $\pm 0.009$& $1.225$ & $\pm 0.008$& $1.113$ & $\pm 0.008$ \\ $y_t$ & $0.8464$ & $\pm 0.0042$& $0.8163$ & $\pm 0.0043$& $0.7765$ & $\pm 0.0044$& $0.7220$ & $\pm 0.0045$ \\ \hline
$\theta_{12}$ & $0.22704$ & $\pm 0.00082$& $0.22705$ & $\pm 0.00082$& $0.22705$ & $\pm 0.00082$& $0.22705$ & $\pm 0.00082$ \\ $\theta_{23}/10^{-2}$ & $4.275$ & $\pm 0.042$& $4.307$ & $\pm 0.042$& $4.339$ & $\pm 0.042$& $4.394$ & $\pm 0.043$ \\ $\theta_{13}/10^{-3}$ & $3.77$ & $\pm 0.08$& $3.80$ & $\pm 0.08$& $3.83$ & $\pm 0.08$& $3.88$ & $\pm 0.08$ \\ $\delta$ & $1.139$ & $\pm 0.023$& $1.139$ & $\pm 0.023$& $1.139$ & $\pm 0.023$& $1.139$ & $\pm 0.023$  \\ 
$J/10^{-5}$ & 3.23 & $\pm0.08$ & 3.28 & $\pm0.08$ & 3.33 & $\pm0.09$ & 3.41 & $\pm0.09$
\\ \hline
$y_e/10^{-6}$ & $2.8279$ & $\pm 0.0007$& $2.8395$ & $\pm 0.0011$& $2.8528$ & $\pm 0.0016$& $2.8662$ & $\pm 0.0024$ \\ $y_{\mu}/10^{-4}$ & $5.9573$ & $\pm 0.0015$& $5.9817$ & $\pm 0.0022$& $6.0099$ & $\pm 0.0034$& $6.0381$ & $\pm 0.0051$ \\ $y_{\tau}/10^{-2}$ & $1.0119$ & $\pm 0.0003$& $1.0161$ & $\pm 0.0004$& $1.0209$ & $\pm 0.0006$& $1.0257$ & $\pm 0.0009$ \\ \hline
$g_3$ & $1.0489$ & $\pm 0.0030$& $1.0047$ & $\pm 0.0026$& $0.9477$ & $\pm 0.0022$& $0.8713$ & $\pm 0.0017$ \\ $g_2$ & $0.63746$ & $\pm 0.00003$& $0.63273$ & $\pm 0.00003$& $0.62582$ & $\pm 0.00003$& $0.61480$ & $\pm 0.00003$ \\ $g_1$ & $0.468064$ & $\pm 0.000027$& $0.470536$ & $\pm 0.000027$& $0.474318$ & $\pm 0.000028$& $0.480829$ & $\pm 0.000029$ \\ \hline
\end{tabular}
}

\caption{Running SM parameters and 1$\sigma$ HPD intervals in the $\overline{\text{DR}}$ renormalization scheme using the \textbf{2024 PDG} data as input. }\label{DR, 2024}
\end{table}
\clearpage

\begin{table}[]
\centering
\renewcommand{\arraystretch}{1.25}
\resizebox{0.95\textwidth}{!}{
 \begin{tabular}{|c|ll|ll|ll|ll|}
\hline
SM Quantity & \multicolumn{2}{c|}{$\tan\beta=5$}& \multicolumn{2}{c|}{$\tan\beta=10$} & \multicolumn{2}{c|}{$\tan\beta=30$} & \multicolumn{2}{c|}{$\tan\beta=50$}\\
\hline$y_u/10^{-6}$ & $2.82$ & $^{+0.61}_{-0.36}$ & $2.79$ & $^{+0.60}_{-0.36}$ & $2.79$ & $^{+0.60}_{-0.36}$ & $2.83$ & $^{+0.61}_{-0.36}$  \\ $y_d/10^{-5}$ & $0.49$ & $^{+0.05}_{-0.02}$ & $0.49$ & $^{+0.05}_{-0.02}$ & $0.52$ & $^{+0.05}_{-0.02}$ & $0.60$ & $^{+0.06}_{-0.03}$  \\ $y_s/10^{-4}$ & $0.99$ & $^{+0.08}_{-0.05}$ & $0.99$ & $^{+0.08}_{-0.05}$ & $1.04$ & $^{+0.08}_{-0.05}$ & $1.20$ & $^{+0.09}_{-0.07}$  \\ $y_c/10^{-3}$ & $1.40$ & $^{+0.05}_{-0.04}$ & $1.39$ & $^{+0.05}_{-0.04}$ & $1.39$ & $^{+0.05}_{-0.04}$ & $1.41$ & $^{+0.05}_{-0.04}$  \\ $y_b/10^{-2}$ & $0.549$ & $^{+0.007}_{-0.006}$ & $0.551$ & $^{+0.008}_{-0.006}$ & $0.601$ & $^{+0.010}_{-0.007}$ & $0.768$ & $^{+0.018}_{-0.015}$  \\ $y_t$ & $0.5076$ & $^{+0.0084}_{-0.0086}$ & $0.4965$ & $^{+0.0078}_{-0.0081}$ & $0.5041$ & $^{+0.0077}_{-0.0088}$ & $0.5340$ & $^{+0.0087}_{-0.0105}$  \\ \hline
$\theta_{12}$ & $0.22702$ & $^{+0.00083}_{-0.00081}$ & $0.22702$ & $^{+0.00083}_{-0.00081}$ & $0.22702$ & $^{+0.00083}_{-0.00081}$ & $0.22702$ & $^{+0.00083}_{-0.00081}$  \\ $\theta_{23}/10^{-2}$ & $3.858$ & $^{+0.038}_{-0.039}$ & $3.869$ & $^{+0.036}_{-0.041}$ & $3.821$ & $^{+0.035}_{-0.041}$ & $3.679$ & $^{+0.037}_{-0.038}$  \\ $\theta_{13}/10^{-3}$ & $3.40$ & $^{+0.07}_{-0.08}$ & $3.41$ & $^{+0.08}_{-0.07}$ & $3.37$ & $^{+0.07}_{-0.07}$ & $3.24$ & $^{+0.08}_{-0.07}$  \\ $\delta$ & $1.139$ & $^{+0.023}_{-0.023}$ & $1.139$ & $^{+0.023}_{-0.023}$ & $1.139$ & $^{+0.023}_{-0.023}$ & $1.139$ & $^{+0.023}_{-0.023}$  \\ 
$J/10^{-5}$ & 2.64 & $^{+0.07}_{-0.07}$ & 2.65 & $^{+0.07}_{-0.07}$ & 2.58 & $^{+0.07}_{-0.07}$ & 2.39 &
$^{+0.06}_{-0.06}$
 \\ \hline
$y_e/10^{-6}$ & $2.0011$ & $^{+0.0007}_{-0.0007}$ & $2.0098$ & $^{+0.0007}_{-0.0009}$ & $2.1128$ & $^{+0.0028}_{-0.0021}$ & $2.4263$ & $^{+0.0123}_{-0.0103}$  \\ $y_{\mu}/10^{-4}$ & $4.2157$ & $^{+0.0016}_{-0.0015}$ & $4.2340$ & $^{+0.0015}_{-0.0019}$ & $4.4516$ & $^{+0.0060}_{-0.0044}$ & $5.1137$ & $^{+0.0260}_{-0.0217}$  \\ $y_{\tau}/10^{-2}$ & $0.7170$ & $^{+0.0003}_{-0.0003}$ & $0.7224$ & $^{+0.0003}_{-0.0003}$ & $0.7891$ & $^{+0.0010}_{-0.0010}$ & $1.0096$ & $^{+0.0059}_{-0.0052}$  \\ \hline
\end{tabular}
}
\caption{GUT-scale values ($M_\textrm{GUT}=2\times 10^{16}$ GeV) with $M_\textrm{SUSY} = \mathbf{3}\,\mathbf{TeV}$ using \textbf{2022 PDG} data as input. 
Here, ``SM Quantities'' $y_f \equiv y_f^\text{SM}$ refer to $y_{u,c,t}^\text{SM}=y_{u,c,t}^\text{MSSM}\sin\beta$, $y_{d,s,b}^\text{SM}=y_{d,s,b}^\text{MSSM}\cos\beta$, $y_{e,\mu,\tau}^\text{SM}=y_{e,\mu,\tau}^\text{MSSM}\cos\beta$, $\theta_{ij}^\text{SM}=\theta_{ij}^\text{MSSM}$, and $\delta^\text{SM}=\delta^\text{MSSM}$; see text for details (especially Eqs.~\eqref{eq:th:01}-\eqref{eq:mixing} for the implementation of SUSY threshold corrections).
}
\label{2022, GUT, 3TeV}

\vspace{5mm}

\renewcommand{\arraystretch}{1.4}
\resizebox{0.95\textwidth}{!}{
 \begin{tabular}{|c|ll|ll|ll|ll|}
\hline
SM Quantity & \multicolumn{2}{c|}{$\tan\beta=5$}& \multicolumn{2}{c|}{$\tan\beta=10$} & \multicolumn{2}{c|}{$\tan\beta=30$} & \multicolumn{2}{c|}{$\tan\beta=50$}\\
\hline$y_u/10^{-6}$ & $2.79$ & $\pm 0.07$& $2.76$ & $\pm 0.07$& $2.76$ & $\pm 0.07$& $2.79$ & $\pm 0.07$ \\ $y_d/10^{-5}$ & $0.49$ & $\pm 0.01$& $0.49$ & $\pm 0.01$& $0.52$ & $\pm 0.01$& $0.59$ & $\pm 0.01$ \\ $y_s/10^{-4}$ & $0.97$ & $\pm 0.01$& $0.98$ & $\pm 0.01$& $1.03$ & $\pm 0.02$& $1.18$ & $\pm 0.02$ \\ $y_c/10^{-3}$ & $1.41$ & $\pm 0.03$& $1.39$ & $\pm 0.03$& $1.39$ & $\pm 0.03$& $1.41$ & $\pm 0.03$ \\ $y_b/10^{-2}$ & $0.548$ & $\pm 0.006$& $0.550$ & $\pm 0.006$& $0.600$ & $\pm 0.007$& $0.766$ & $\pm 0.013$ \\ $y_t$ & $0.5058$ & $\pm 0.0084$& $0.4947$ & $\pm 0.0079$& $0.5020$ & $\pm 0.0081$& $0.5312$ & $\pm 0.0095$ \\ \hline
$\theta_{12}$ & $0.22702$ & $\pm 0.00082$& $0.22702$ & $\pm 0.00082$& $0.22702$ & $\pm 0.00082$& $0.22702$ & $\pm 0.00082$ \\ $\theta_{23}/10^{-2}$ & $3.862$ & $\pm 0.038$& $3.874$ & $\pm 0.038$& $3.823$ & $\pm 0.038$& $3.679$ & $\pm 0.037$ \\ $\theta_{13}/10^{-3}$ & $3.40$ & $\pm 0.07$& $3.41$ & $\pm 0.07$& $3.37$ & $\pm 0.07$& $3.24$ & $\pm 0.07$ \\ $\delta$ & $1.139$ & $\pm 0.023$& $1.139$ & $\pm 0.023$& $1.139$ & $\pm 0.023$& $1.139$ & $\pm 0.023$  \\ 
$J/10^{-5}$ & 2.64 & $\pm 0.07$ & $2.65$ & $\pm0.07$ & 2.58 & $\pm0.07$ & 2.39 &
$\pm0.06$\\ \hline
$y_e/10^{-6}$ & $2.0010$ & $\pm 0.0007$& $2.0095$ & $\pm 0.0008$& $2.1128$ & $\pm 0.0020$& $2.4248$ & $\pm 0.0087$ \\ $y_{\mu}/10^{-4}$ & $4.2154$ & $\pm 0.0016$& $4.2334$ & $\pm 0.0017$& $4.4515$ & $\pm 0.0042$& $5.1104$ & $\pm 0.0183$ \\ $y_{\tau}/10^{-2}$ & $0.7170$ & $\pm 0.0003$& $0.7223$ & $\pm 0.0003$& $0.7889$ & $\pm 0.0008$& $1.0089$ & $\pm 0.0042$ \\ \hline
\end{tabular}
}
\caption{GUT-scale values ($M_\textrm{GUT}=2\times 10^{16}$ GeV) with $M_\textrm{SUSY} = \mathbf{3}\,\mathbf{TeV}$ using \textbf{2024 PDG} data as input. For the definition of ``SM Quantities'' cf.\ Table \ref{2022, GUT, 3TeV}.
}
\label{2024, GUT, 3TeV}
\end{table}

\begin{table}[]
\centering
\renewcommand{\arraystretch}{1.25}
\resizebox{0.95\textwidth}{!}{
 \begin{tabular}{|c|ll|ll|ll|ll|}
\hline
SM Quantity & \multicolumn{2}{c|}{$\tan\beta=5$}& \multicolumn{2}{c|}{$\tan\beta=10$} & \multicolumn{2}{c|}{$\tan\beta=30$} & \multicolumn{2}{c|}{$\tan\beta=50$}\\
\hline$y_u/10^{-6}$ & $2.83$ & $^{+0.61}_{-0.37}$ & $2.80$ & $^{+0.60}_{-0.36}$ & $2.80$ & $^{+0.60}_{-0.36}$ & $2.83$ & $^{+0.61}_{-0.37}$  \\ $y_d/10^{-5}$ & $0.50$ & $^{+0.05}_{-0.02}$ & $0.51$ & $^{+0.05}_{-0.02}$ & $0.53$ & $^{+0.05}_{-0.02}$ & $0.60$ & $^{+0.06}_{-0.03}$  \\ $y_s/10^{-4}$ & $1.02$ & $^{+0.08}_{-0.05}$ & $1.02$ & $^{+0.08}_{-0.05}$ & $1.07$ & $^{+0.08}_{-0.05}$ & $1.21$ & $^{+0.09}_{-0.07}$  \\ $y_c/10^{-3}$ & $1.41$ & $^{+0.05}_{-0.04}$ & $1.39$ & $^{+0.05}_{-0.04}$ & $1.39$ & $^{+0.05}_{-0.04}$ & $1.41$ & $^{+0.05}_{-0.04}$  \\ $y_b/10^{-2}$ & $0.555$ & $^{+0.008}_{-0.006}$ & $0.557$ & $^{+0.008}_{-0.006}$ & $0.603$ & $^{+0.010}_{-0.007}$ & $0.747$ & $^{+0.016}_{-0.014}$  \\ $y_t$ & $0.4996$ & $^{+0.0077}_{-0.0081}$ & $0.4902$ & $^{+0.0073}_{-0.0077}$ & $0.4966$ & $^{+0.0071}_{-0.0083}$ & $0.5205$ & $^{+0.0078}_{-0.0096}$  \\ \hline
$\theta_{12}$ & $0.22702$ & $^{+0.00083}_{-0.00081}$ & $0.22702$ & $^{+0.00083}_{-0.00081}$ & $0.22702$ & $^{+0.00083}_{-0.00081}$ & $0.22702$ & $^{+0.00083}_{-0.00081}$  \\ $\theta_{23}/10^{-2}$ & $3.891$ & $^{+0.038}_{-0.039}$ & $3.900$ & $^{+0.039}_{-0.038}$ & $3.856$ & $^{+0.039}_{-0.038}$ & $3.737$ & $^{+0.034}_{-0.041}$  \\ $\theta_{13}/10^{-3}$ & $3.43$ & $^{+0.08}_{-0.07}$ & $3.44$ & $^{+0.08}_{-0.07}$ & $3.40$ & $^{+0.07}_{-0.07}$ & $3.29$ & $^{+0.08}_{-0.07}$  \\ $\delta$ & $1.139$ & $^{+0.023}_{-0.023}$ & $1.139$ & $^{+0.023}_{-0.023}$ & $1.139$ & $^{+0.023}_{-0.023}$ & $1.139$ & $^{+0.023}_{-0.023}$  \\ 
$J/10^{-5}$ & 2.70 & $^{+0.07}_{-0.07}$ & 2.72 & $^{+0.07}_{-0.07}$ & 2.65 & $^{+0.07}_{-0.07}$ & 2.49 & $^{+0.07}_{-0.07}$ 
\\ \hline
$y_e/10^{-6}$ & $2.0495$ & $^{+0.0011}_{-0.0011}$ & $2.0576$ & $^{+0.0012}_{-0.0011}$ & $2.1547$ & $^{+0.0023}_{-0.0028}$ & $2.4357$ & $^{+0.0107}_{-0.0091}$  \\ $y_{\mu}/10^{-4}$ & $4.3175$ & $^{+0.0022}_{-0.0024}$ & $4.3347$ & $^{+0.0024}_{-0.0024}$ & $4.5398$ & $^{+0.0049}_{-0.0059}$ & $5.1334$ & $^{+0.0224}_{-0.0192}$  \\ $y_{\tau}/10^{-2}$ & $0.7342$ & $^{+0.0004}_{-0.0003}$ & $0.7396$ & $^{+0.0004}_{-0.0004}$ & $0.8038$ & $^{+0.0009}_{-0.0011}$ & $1.0067$ & $^{+0.0052}_{-0.0044}$  \\ \hline
\end{tabular}
}
\caption{GUT-scale values ($M_\textrm{GUT}=2\times 10^{16}$ GeV) with $M_\textrm{SUSY} = \mathbf{10}\,\mathbf{TeV}$ using \textbf{2022 PDG} data as input. For the definition of ``SM Quantities'' cf.\ Table \ref{2022, GUT, 3TeV}.}
\label{2022, GUT, 10TeV}

\vspace{5mm}

\renewcommand{\arraystretch}{1.4}
\resizebox{0.95\textwidth}{!}{
 \begin{tabular}{|c|ll|ll|ll|ll|}
\hline
SM Quantity & \multicolumn{2}{c|}{$\tan\beta=5$}& \multicolumn{2}{c|}{$\tan\beta=10$} & \multicolumn{2}{c|}{$\tan\beta=30$} & \multicolumn{2}{c|}{$\tan\beta=50$}\\
\hline$y_u/10^{-6}$ & $2.79$ & $\pm 0.07$& $2.77$ & $\pm 0.07$& $2.77$ & $\pm 0.07$& $2.79$ & $\pm 0.07$ \\ $y_d/10^{-5}$ & $0.50$ & $\pm 0.01$& $0.50$ & $\pm 0.01$& $0.53$ & $\pm 0.01$& $0.60$ & $\pm 0.01$ \\ $y_s/10^{-4}$ & $1.00$ & $\pm 0.01$& $1.00$ & $\pm 0.01$& $1.05$ & $\pm 0.02$& $1.19$ & $\pm 0.02$ \\ $y_c/10^{-3}$ & $1.41$ & $\pm 0.03$& $1.40$ & $\pm 0.03$& $1.40$ & $\pm 0.03$& $1.41$ & $\pm 0.03$ \\ $y_b/10^{-2}$ & $0.554$ & $\pm 0.006$& $0.556$ & $\pm 0.006$& $0.602$ & $\pm 0.007$& $0.745$ & $\pm 0.012$ \\ $y_t$ & $0.4979$ & $\pm 0.0078$& $0.4885$ & $\pm 0.0074$& $0.4945$ & $\pm 0.0076$& $0.5179$ & $\pm 0.0086$ \\ \hline
$\theta_{12}$ & $0.22702$ & $\pm 0.00082$& $0.22702$ & $\pm 0.00082$& $0.22702$ & $\pm 0.00082$& $0.22702$ & $\pm 0.00082$ \\ $\theta_{23}/10^{-2}$ & $3.896$ & $\pm 0.039$& $3.905$ & $\pm 0.039$& $3.860$ & $\pm 0.038$& $3.735$ & $\pm 0.038$ \\ $\theta_{13}/10^{-3}$ & $3.43$ & $\pm 0.07$& $3.44$ & $\pm 0.07$& $3.40$ & $\pm 0.07$& $3.29$ & $\pm 0.07$ \\ $\delta$ & $1.139$ & $\pm 0.023$& $1.139$ & $\pm 0.023$& $1.139$ & $\pm 0.023$& $1.139$ & $\pm 0.023$  \\ 
$J/10^{-5}$ & 2.70 & $\pm0.07$ & 2.72 & $\pm0.07$ & 2.65 & $\pm 0.07$ & 2.49 &
$\pm0.07$\\ \hline
$y_e/10^{-6}$ & $2.0493$ & $\pm 0.0011$& $2.0573$ & $\pm 0.0011$& $2.1541$ & $\pm 0.0022$& $2.4341$ & $\pm 0.0078$ \\ $y_{\mu}/10^{-4}$ & $4.3171$ & $\pm 0.0023$& $4.3341$ & $\pm 0.0024$& $4.5386$ & $\pm 0.0046$& $5.1300$ & $\pm 0.0164$ \\ $y_{\tau}/10^{-2}$ & $0.7343$ & $\pm 0.0004$& $0.7395$ & $\pm 0.0004$& $0.8036$ & $\pm 0.0009$& $1.0059$ & $\pm 0.0038$ \\ \hline
\end{tabular}
}
\caption{GUT-scale values ($M_\textrm{GUT}=2\times 10^{16}$ GeV) with $M_\textrm{SUSY} = \mathbf{10}\,\mathbf{TeV}$ using \textbf{2024 PDG} data as input. For the definition of ``SM Quantities'' cf.\ Table \ref{2022, GUT, 3TeV}.}
\label{2024, GUT, 10TeV}
\end{table}


\bibliographystyle{style}
\bibliography{reference}
\end{document}